# King of the Hill: C2 for Next Generation Swarm Warfare

Takuma Adams
*Defence Science and Technology Group*
takuma.adams@defence.gov.au

Timothy McLennan-Smith
*Defence Science and Technology Group*
timothy.mclennan-smith@defence.gov.au

**Abstract**

As the reliability of cheap, off-the-shelf autonomous platforms increases, so does the risk posed by intelligent multi-agent systems to military operations. In the contemporary context of the Russo-Ukrainian war alone, we have seen autonomous aerial vehicles and surface vessels deployed both individually and in multitude to deliver critical effects to both sides. While there is a large body of literature on tactical level communications and interactions between agents, the exploration of high-level command and control (C2) structures that will underpin future autonomous multi-agent military operations is a less explored area of research. We propose a quantitative game-theoretic framework to study effective C2 structures in cooperative and competitive multi-agent swarming scenarios. To test our framework, we construct a virtual environment where two adversarial swarms compete to achieve outcomes comparable to real-world scenarios. The framework we present in this paper enables us to quickly test and interrogate different C2 configurations in multi-agent systems to explore C2 as a force multiplier when at a force disadvantage.

## 1 INTRODUCTION

Military conflicts in the 21st century are increasingly seeing the use of low-cost autonomous platforms to provide significant asymmetric effects [1-3]. The utilization of uncrewed aerial systems (UAS) in the recent Russo-Ukrainian conflict is seeing a spur in the development of autonomous capabilities [1, 2]. While the impact single platforms can achieve is demonstrably significant, platforms exhibiting multi-agent behavior (e.g. swarming or flocking) have the potential to deliver greater military results.

The swarming [6-9] and self-synchronizing [10-15] behaviors critical to this new generation of platforms are frequently observed across many different natural and cyber-physical systems. While the study of such systems is well established in the literature, they are usually explored in isolation. Contrary to this usual separation, it can be argued that collective behavior cannot occur independently of synchronization. That is, the communication and coordination between agents, be they heterogeneous or homogenous, form a key aspect of the underlying dynamics that drive such systems.

Early work by Olfati-Saber sees the uncoupled interplay of external spatial states and internal synchronization dynamics of agents in the system. This concept is further explored with the '*swarmalator*' [3, 4] where the spatial and synchronization dynamics of each agent in the system are coupled. In this model, the internal decision state of the agent leads to collective synchronization which drives the spatial dynamics of agents. Swarmalators are an active area of research with applications in swarm robotics [5, 6] in addition to other cyber-physical systems.

The concept of the swarmalator was then extended to a competitive model by McLennan-Smith et al. [6] by introducing two population sets to study emerging adversarial behaviors in the system. The phase transitions observed by the two subgroups of agents were likened to and analyzed through a military maneuvers lens. The adversarial behavior observed in the extended swarmalator model lends itself to the study of how one population can employ a strategy to outperform its adversary.

We differentiate ourselves from previous work by introducing a game-theoretic lens to study the adversarial behavior of swarms. Further, we employ a hierarchical command and control (C2) structure to explore the emergence of swarm intelligence in the context of competing heterogeneous (human and machine) teams. Using this framework, we explore C2 as a force multiplier and the dominant strategies each population employs to achieve their conflicting goals within the limitations of their decision space.

This paper makes the following contributions:

1. Layering of swarmalator dynamics with game-theoretic drivers to demonstrate enhanced decision dynamics

2. Development of a heterogeneous multi-agent

swarming environment with hierarchical C2.

The remainder of the paper is organized as follows. Sections 2 and 0 provide a detailed overview of the Swarmalator model and our environment. This is followed by a discussion of the game-theoretic component in Section 0 before final conclusions are drawn.

## 2 SETTING THE SCENE

We consider a fictitious scenario with two adversaries, or players, labeled Blue and Red. Each player uses an autonomous swarm to secure and maintain a presence in a patch of land much like in the children's game '*king of the hill*[1]'. The swarms achieve this by entering the 'hill' while simultaneously repelling adversarial forces. The swarm's C2 can be likened to a point to multi-point or all-to-all structure with distance-based attenuation with no hierarchy between agents. To this end, both autonomous swarms are attempting to gain both a spatial and decision advantage. It is worth noting that since we do not restrict the number of spatial degrees-of-freedom in our work, the autonomous agents move how you might imagine a quad-copter, or conventional rotor-wing UAS, to move.

Supporting the autonomous swarms, both players maintain a headquarters staffed by humans organized in a generic C2 hierarchy. In a real-world context, this can be likened to a headquarters' staff conducting an autonomous swarm operation. While no spatial dynamics apply, both headquarters still try to gain a decision advantage over their adversary to further their operational advantage. A single operator within the headquarters then influences the decision dynamics of the swarm as shown in Figure 1.

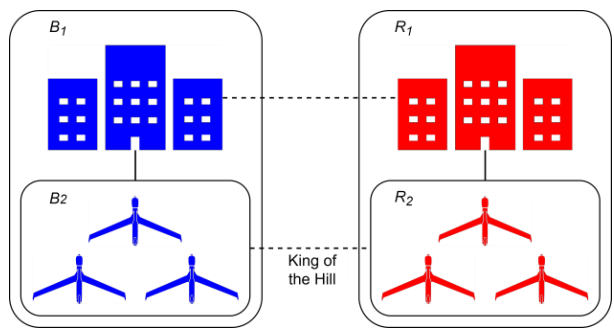

*Figure 1: Illustration of scenario. Solid lines indicate hierarchical C2 linkages while dashed lines represent adversarial interactions.*

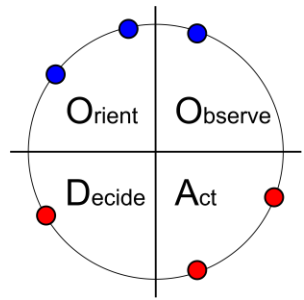

*Figure 2: Representation of OODA loop projected onto the unit circle. The blue and red circles depict the internal decision states of different agents.*

## 3 THE MODEL

The decision-making component of our model leverages Kuramoto dynamics [7] to represent the synchronizing behavior of swarms and autonomous systems [8, 9, 10]. While many different decision-making frameworks exist [18-20], we leverage the established mathematical definition of Boyd's Observe, Orient, Decide, Act (OODA) loop [11]. That is, we use the OODA loop to project an agent's internal decision state (phase) to the unit circle as depicted in Figure 2. Our model's implementation of OODA paired with Kuramoto dynamics to capture decision-making elements of command and control has been validated for human decision-making in military contexts [22-25].

We formally represent the two players as sets of decision-making agents, labeled Blue and Red. These two sets are further partitioned to represent the hierarchical C2 component of our model. Therefore, we get four sets of agents denoted $B_1$, $B_2$ and $R_1$, $R_2$ for the headquarters and swarm subsets of the Blue and Red players respectively as depicted in Figure 1.

We let $L$ be the total number of agents in the system. For this paper, we set both headquarters to contain 21 agents and the blue and red swarms to contain 20 and 25 agents respectively for a total of $L = 86$ agents. Thus in the absence of superior decision-making, the scenario will favor Red. The precise network structure for the two players is illustrated in Figure 3, which consists of two hierarchal Headquarters that interact with each other at the lowest ranking levels of the hierarchy. In these headquarters, one of the lowest ranking agents is designated as the 'HQ swarm controllers' and is responsible for sending instructions to their collective swarm which impacts the overall decision states. For the sake of notation, we denote the set of decision-making agents in the headquarters and swarm agents as $S_1 = B_1 \cup R_1$ and $S_2 = B_2 \cup R_2$ such that

$$S_1 = \{\theta \in Headquarters\}, S_2 = \{\theta \in Swarm\}.$$

---

[1] The game's objective is for a single player to remain on the 'hill' for as long as possible. Adversaries compete to take their place.

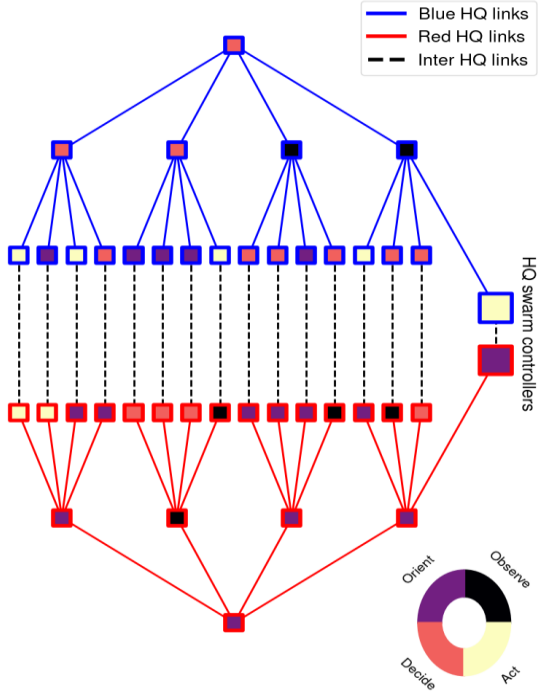

Figure 3: C2 structures and initial OODA states of the Blue and Red HQs at the start of the C2 Game.

As with previous work [13-15], we define the swarmalator dynamics in terms of time varying decision states, θ, and positions, $x$. The coupled system of equations is given by

$$\dot{\theta}_i = \omega_i + \sigma \sum_{j=1}^{L} \frac{A_{ij}(x)}{d_i^{\beta_1} d_j^{\beta_2}} \sin(N_{ij}\theta_j - N_{ij}^T \theta_i + \Phi_{ij}), \quad (1)$$

$$\dot{x}_i = F_i^{att}(x,\theta) - F_i^{rep}(x) + F_i^{field}(x,\theta), \quad (2)$$

for $i \in \{1,2,\dots,L\}$ where $\omega_i$ is the natural frequency (decision speed) and $d_i$ is the degree of the $i$th agent while $\sigma$ denotes the coupling strength between agents. The connections between agents are encoded in $A$. The frustration term, $\Phi_{ij}$, controls whether $i$th agent is trying to lag or lead the $j$th agent's decision cycle.

As initially explored by Kalloniatis [12], we leverage the abstraction of agents to phase oscillators to fully characterize the type of agent by their intrinsic decision speed $\omega$. This allows us to encode the heterogeneous nature of our system—with humans and the swarms—in the matrix $N$. We assume autonomous agents can make decisions faster than humans, hence represent the decision frequency of an agent as a multiple of the human's decision speed such that

$$\omega_i^{(Swarm)} = N_{ij}\omega_j^{(Headquaters)},$$

$$\theta_i^{(Swarm)} = N_{ij}\theta_j^{(Headquaters)},$$

for some $N_{ij} > 1$.

Finally, the spatial dynamics for the $i$th agent are governed by $F_i^{att}$, $F_i^{rep}$, and $F_i^{field}$ which are given as functions of the agent's current position and phase. Their functional forms and intuition are provided below.

$$F_i^{att}(x,\theta) = O_i^{Swarm} \sum_{j \in S^2} \frac{(1 + \alpha_{ij}(x)\cos(\theta_j - \theta_i + \Phi_{ij}^S))}{\sqrt{(1 + |x_j - x_i|^2)}}$$
$$\times \frac{x_j - x_i}{|x_j - x_i|}, \quad (3)$$

$$F_i^{rep}(x) = \sum_{j \in S^2} \frac{\rho}{(1 + |x_j - x_i|^2)^2} \cdot \frac{x_j - x_i}{|x_j - x_i|}, \quad (4)$$

$$F_i^{field}(x,\theta) = -c^2 O_i^{Headquaters} x_i 1_{\{|x_i|>1\}},$$

$$\alpha_{ij}(x) = \alpha_{ij}\left(1_{\{|x_i|<1\}}\left(1 - 1_{\{|x_j|>1\}}\frac{c_3|x_j|}{1 + c_3|x_j|}\right)\right). \quad (5)$$

$F_i^{att}$ is dependent on the synchronization of the swarm and controls the $i$th agent's ability to move towards the center of the hill. These terms additionally capture the following rules:

- Swarms of a similar population will move towards one another if they close in their decision state.
- Swarm agents will engage with adversarial agents if they are in, or within proximity to, the goal area of 'king of the hill'. Agents that are ahead by an OODA state can engage in these dynamics with full effectiveness.
- The strength of attraction between swarm agents diminishes the further apart they are.
- The collective decision-making synchronization of each swarm affects its ability to engage in these dynamics.

$F_i^{rep}$ captures the repulsion between agents to prevent collisions.

$F_i^{field}$ captures the movement of the swarming agents in response to the C2 Game and in this scenario forces agents towards the hill. As such this term is only active when the swarming agent is outside of the patch of interest. Here the ability of the swarm to engage in the rules of this game is dependent upon the collective decision-making synchronization of the respective headquarters.

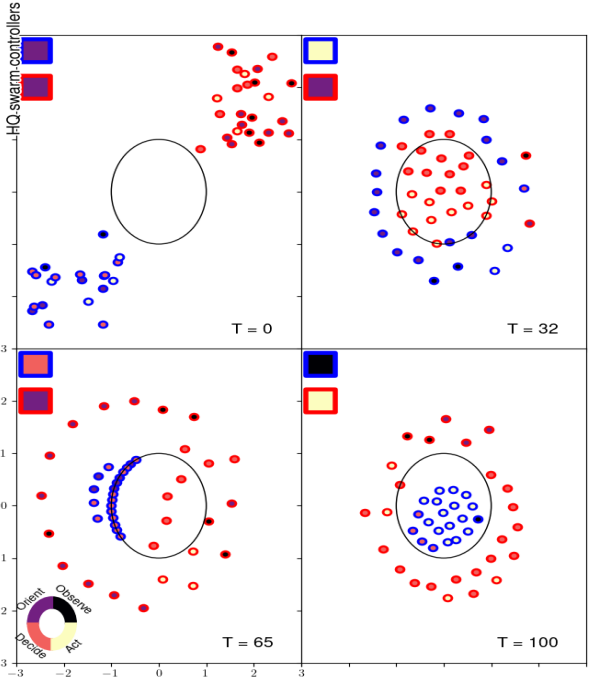

*Figure 4: Snapshots of swarm states for the two players at four time intervals over the course of a game of 'king of the hill'. Here $\Phi = 0.5\,\pi$ (Blue intends to be one OODA state ahead of Red). The decision state of the HQ swarm controllers have been superimposed on the top left.*

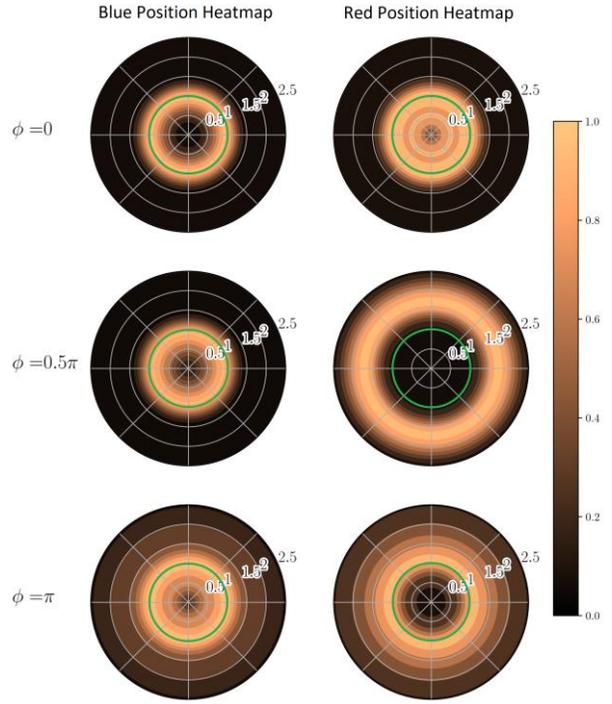

*Figure 5: Heatmaps denoting the probability density for the Blue and Red swarming agent locations over the C2 Game. Repeated for various values of Blue frustration: $\Phi = 0$ (Blue intends to match Red's OODA state), $\Phi = 0.5\,\pi$ (Blue intends to be one OODA state ahead of Red), $\Phi = \pi$ (Blue intends to be two OODA states ahead of Red). The green circle represents the 'hill'.*

We provide an overview of the synchronization and spatial behaviors of our environment below. Figure 4 illustrates the evolution of spatial dynamics over a single iteration of the game. In this set-up, Blue intends to be one OODA state ahead of Red and we can see their intended decision advantage payoff at the end of the game where all Blue swarming agents occupy the hill. We also investigate the impact of Blue's frustration on spatial dynamics. The difference frustration can make to the game outcomes is demonstrated in Figure 5 where we plot position heatmaps for the blue and red swarms over the course of the C2 Game for different values of the Blue frustration. As Red has the intrinsic advantage of a larger swarm, when $\Phi = 0$ Blue is generally located on the perimeter of the circle whilst Red stays inside. With frustrations of $\Phi = 0.5\pi$ and $\Phi = \pi$, Blue is able to make up for its small size through superior decision-making as the Blue swarm is generally located within the circle. However, it should be noted that this success is diminished for $\Phi = \pi$ as trying to be too far ahead in the decision-making cycle can impede success in this model.

An important addition of this paper is the improvement of the scaling term for synchronization dynamics. In previous works, the effect of synchronizing was often scaled by either a constant term $1/L$ or by the degree of the $i$th agent. Instead, we apply research [13] that shows the benefit of scaling by the degree of both agents. The implications of this new scaling term is particularly relevant to heterogeneous systems where a human may not be able to simultaneously interact with as many agents as an autonomous agent can. The scaling term is given by

$$\frac{A_{ij}(x)}{d_i^{\beta_i^1} d_j^{\beta_j^2}},$$

for $i, j \in \{1, 2, \ldots, L\}$ where $\beta_i^1 = 1$, $\beta_j^2 = 0$ if $i \in S_2$ and $\beta_j^2 = 1$ if $i \in S_1$.

That is, humans are inhibited by the number of connections while autonomous agents are not. In addition, communication between swarming agents in the field is subject to distance-based attenuation. While a simple model, this attenuation is captured in the below equation

$$A_{ij}(x) = \begin{cases} A_{ij} & , \text{if } i,j \in S_1 \\ \dfrac{A_{ij}}{1 + c^1|x_j - x_i|} & , \text{if } i,j \in S_2 \end{cases}. \quad (6)$$

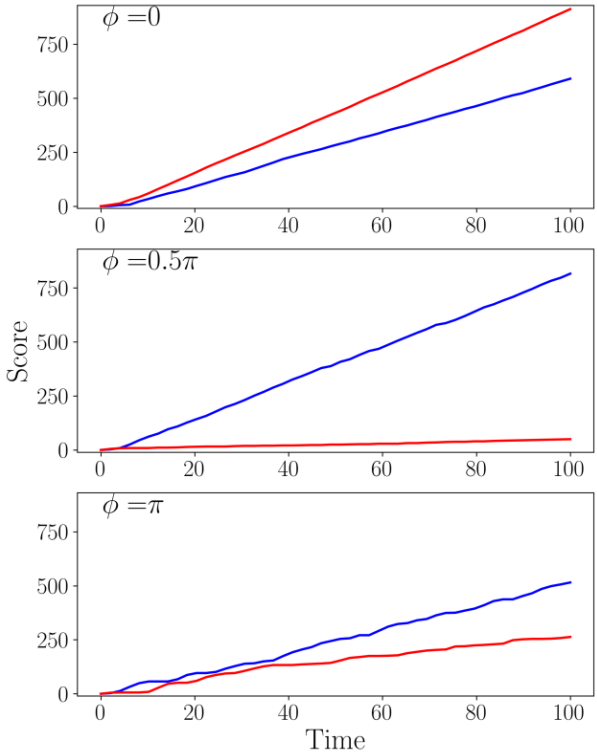

Figure 6: Blue and Red scores over the course of the 'king of the hill' for various levels of Blue frustration.

The success of a player is quantified by both the duration and number of swarming agents occupying the hill. The score of the Blue and Red players at time $t$ are given by

$$\Omega_B(t) = \int_0^t \mathbf{1}_{\|x_i(\tau)\|_2 \leq 1}, \quad \Omega_R(t) = \int_0^t \mathbf{1}_{\|x_i(\tau)\|_2 \leq 1}, \quad (7)$$

where $\mathbf{1}$ denotes the indicator function and $\|\cdot\|_2$ is the L2 norm. In this scoring, we award a player a 1 point per unit time a swarming agent spends within the hill. Using this scoring system, the quantification of Blue's relative advantage over Red despite their disadvantaged numbers is visualized in Figure 6. Here we plot the scores of the Blue and Red swarms as defined above for various levels of Blue frustration. These scores verify what we observed from the heatmaps in Figure 5, where Blue loses at $\Phi = 0$ due to inferior numbers while winning at $\Phi = 0.5\,\pi$ and $\Phi = \pi$ due to superior decision-making.

## 4    GAME-THEORETIC LAYER

This brings us to introduce a game-theoretic lens to study the influence of decision variables on the dynamics of the decision-making agents. The adversarial decision-making element of the model, with the decision advantage quantified by the 'king of the hill' objective, can be formulated as an extensive form game. In this context, the Blue and Red players control their respective sets of decision-making agents and can move simultaneously. To this end, we can classify the C2 Game presented in this paper as a two-player, zero-sum game subject to the likes of dominant strategies and Nash equilibrium [14].

**Definition 1** (C2 Game): *Let the set of two players $\mathcal{P} = \{Blue, Red\}$ have respective discrete strategies $(\mathcal{S}_B, \mathcal{S}_R)$ with utilities $(\mathcal{U}_B, \mathcal{U}_R)$ defined by Equation 11. The two-player, zero-sum strategic game induced by this configuration is given by the tuple:*

$$\mathcal{G}_{C2} = \langle \mathcal{P}, (\mathcal{S}_B, \mathcal{S}_R), (\mathcal{U}_B, \mathcal{U}_R) \rangle.$$

We assume both players choose actions to gain a decision advantage over their adversary, driving the synchronization dynamics which in turn drives the spatial dynamics. Thus each player decides whether they will lead or lag their opponent to gain a decision advantage by selecting an appropriate frustration $\Phi \in [0, \pi]$. When interpreted through an OODA loop perspective, the frustration term translates to a player's ability to outpace or 'get inside' their adversary's decision-cycle.

The adversarial swarmalator model is run over a finite time horizon of $[0, T_f]$. To implement this game-theoretic layer, we uniformly partition the time horizon into $K$ sets $\{[T_0, T_1), [T_1, T_2), \ldots, [T_{K-1}, T_K]\}$ such that each is long enough to exhibit a meaningful evolution in the dynamics. We define $K$ to equal the number of decision points in the game. Hence, over the time horizon $[0, T_f]$ each player decides on a sequence of $K$ actions yielding a strategy vector

$$\mathcal{S}_B = \left\{\Phi_0^{(B)}, \Phi_1^{(B)}, \ldots, \Phi_K^{(B)}\right\}, \mathcal{S}_R = \left\{\Phi_0^{(R)}, \Phi_1^{(R)}, \ldots, \Phi_K^{(R)}\right\}.$$

For this paper, we consider a discrete action space.

As mentioned previously, the relative success of a given player is quantified by the duration they occupy the 'hill' in the C2 Game. Through modification of Equation 7, we defined the score of a player after the $k$th turn as

$$\Omega_B(k) = \sum_{i \in B_2} \mathbf{1}_{\|x_i(\tau)\|_2 \leq 1}, \quad \Omega_R(k) = \sum_{i \in R_2} \mathbf{1}_{\|x_i(\tau)\|_2 \leq 1}. \quad (8)$$

Thus the relative advantage becomes

$$Q_B(k) = \Omega_B(k) - \Omega_R(k) = -Q_R(k). \quad (9)$$

Using this, we define the utility as

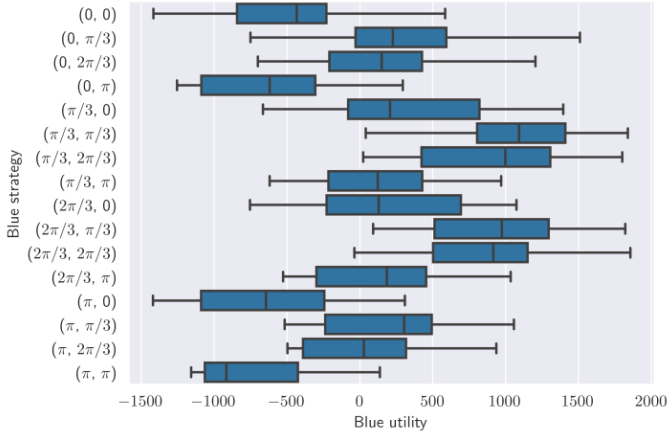

Figure 7: Distribution of Blue utilities for different Blue strategies ordered by first choice in frustration.

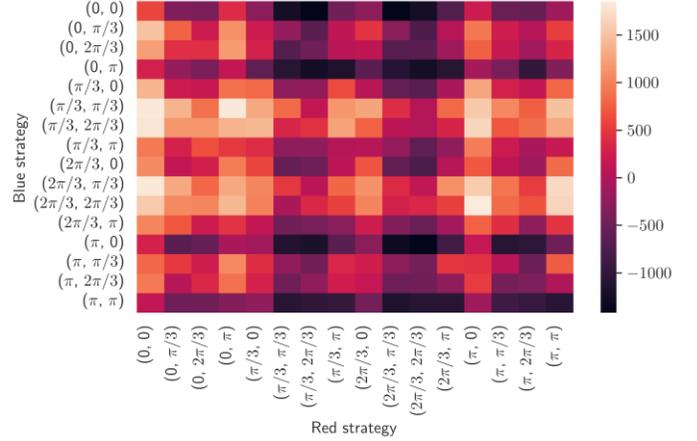

Figure 8: Heapmap of Blue utility for different strategy combinations. Axis labels indicate strategy vectors. Red utility is the inversion of Blue utility due to the zero-sum nature of the game.

$$\mathcal{U}_B(\mathcal{S}_B, \mathcal{S}_R) = \sum_{k=1}^{K} Q_B(k) , \quad (10)$$

$$\mathcal{U}_R(\mathcal{S}_B, \mathcal{S}_R) = \sum_{k=1}^{K} Q_R(k) = -\mathcal{U}_B(\mathcal{S}_B, \mathcal{S}_R) . \quad (11)$$

Consider two strategy vectors $\mathcal{S}_B^*$ and $\mathcal{S}_B'$ for Blue. $\mathcal{S}_B^*$ is said to dominate $\mathcal{S}_B'$ if:

$$\mathcal{U}_B(\mathcal{S}_B^*, \mathcal{S}_R) \geq \mathcal{U}_B(\mathcal{S}_B', \mathcal{S}_R) ,$$

for all $\mathcal{S}_R$. If this is the case for all $\mathcal{S}_B'$, we can say that $\mathcal{S}_B^*$ is dominant. The reverse applies for Red.

### 4.1 Experimental Setup

We consider a discrete action space with four possible choices of frustration for both players. That is both $\Phi^{(B)}$ and $\Phi^{(R)}$ are drawn from $\{0, \pi/3, 2\pi/3, \pi\}$. We then allow the players to re-choose their frustration halfway through the game. Hence, we partition the time horizon into two yielding strategy vectors of:

$$\mathcal{S}_B = \left\{\Phi_0^{(B)}, \Phi_1^{(B)}\right\} , \; \mathcal{S}_R = \left\{\Phi_0^{(R)}, \Phi_1^{(R)}\right\}.$$

Recall we have set both headquarters to contain 21 agents and the blue and red swarms to contain 20 and 25 agents respectively for a total of $L = 86$ decision-making agents. To capture the heterogeneous nature of the humans staffing the headquarters and the autonomous swarming agents, we draw the intrinsic decision speeds for the two populations from different distributions:

$$\omega^{Headquarters} \sim U(0.25, 0.5) , \; \omega^{Swarm} \sim U(1, 2) ,$$

where $U$ is the uniform distribution. The remainder of the parameters for the experimental set-up are provided in Appendix A. The system of differential equations defined by Equations 1 and 2 are then solved using SciPy's solve_ivp with identical initial conditions (aside from the choice of frustration) for each game solve.

### 4.2 Results

We have four actions and two time steps yielding a total of 256 strategy combinations for both players. Given the relatively small search space, we opted to conduct an exhaustive search and play out each of the combinations. To obtain preliminary results and validate the game-theoretic nature of the game we ran a single iteration over all 256 strategy combinations.

In computing the payoffs for each scenario (given by Equation 11), we can see the emergence of dominating strategies for the two players. From Figure 7 we can see that a Blue strategy of $(\pi/3, \pi/3)$ yields the highest utility despite a force disadvantage while $(\pi, \pi)$ results in the worst outcome for Blue. Note that Red's utility is simply a reflection of Figure 7 about the x-axis due to the zero-sum nature of the game. In general, Blue performs better when the player selects strategies with frustrations of $\pi/3$ or $2\pi/3$. These frustrations are interpreted as indenting a single or three OODA state lead Red 1 or 2 in the plots.

While we only experimented over a single iteration of initial values, these results are in agreement with the formulation of the attraction portion of the spatial dynamics given in Equation 2. We can see this decision advantage reflected against Blue in Figure 8 where Blue sees a decrease in performance when Red selects a strategy with frustrations of $\pi/3$ or $2\pi/3$. Interestingly, we also see that Blue generally performs worse when Red chooses a frustration of $\pi$—an intended lead of two OODA states—in the second stage in the game. Conversely Red on average performs better when Blue selects a frustration of $\pi$.

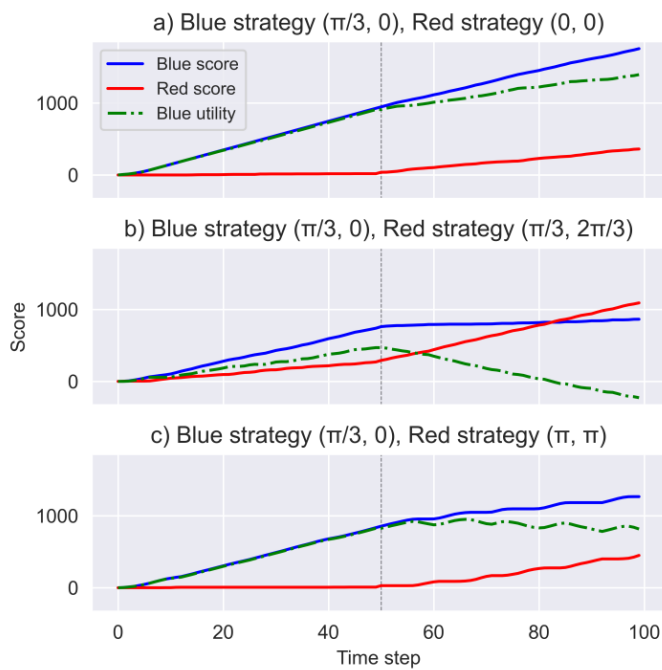

*Figure 9: Different game outcomes for a Blue strategy of $(\pi/3, \pi/3)$. Both players change their strategy midway through the game denoted by the vertical grey line.*

Blue loses more frequently than Red (greater number of negative utilities) due to asymmetric force disadvantage in our experimental setup. Despite this, there are obvious symmetries in game outcomes indicating spatial dynamics are heavily influenced by the selection of decision variables. Without further experimentation it is not possible to solve for Nash Equilibrium.

Finally, we can drill down in the game dynamics for a particular Blue strategy as illustrated in Figure 9 to see C2 acting as a force multiplier. As discussed above, we again see in Figure 9a that Blue significantly outperforms Red when Blue intends to be approximately one OODA state ahead of Red. However, this decision advantage is eroded when both players select the same frustration in the latter half of the game likely due to Red's innate force advantage with more swarming agents. In Figure 9b, we see Red's decision advantage act as a force multiplier wherein being ahead of Blue's decision phase allows them to perform a counter-offensive and expel Blue from the hill. A similar effect of force multiplication is seen in Figure 9c.

## 5 DISCUSSION AND CONCLUSION

The model we presented demonstrates a promising research direction to study the impact of C2 on multi-agent adversarial decision-making systems. Specifically, by extending the swarmalator [3, 4] in an adversarial context [6] to include heterogeneous agents we can study the interplay of agents with different decision-making speeds. Moreover, the added complexity of the system offered through the addition of the hierarchical C2 layer allows us to gain better insights into real-world scenarios.

Preliminary results from our experiments show that there are clear dominant strategies for players in these complex cyber-physical systems. The OODA lens applied to these results makes them intuitive to understand. Further, our results indicate that a C2 advantage acts as a significant force multiplier in multi-agent swarming scenarios which can be validated with real-world data. While additional work is required to fully uncover what effective multi-agent C2 structures look like, we have seen players can win adversarial games despite having a force disadvantage through careful selection of their C2 decision intent.

The C2 Game explored in this paper is a simple vignette that can be abstracted to better understand collaborative and adversarial multi-agent operations. For example, the framework we introduced can easily be applied to counter UAS context where Blue aims to prevent Red agents from accessing restricted airspace (the hill) or where two adversarial swarms attempt to secure a real-world 'hill'. The variety of applications this framework can be adapted to is important since many modern autonomous platforms with swarming capabilities are fit for kinetic, non-kinetic, and intelligence, surveillance, and reconnaissance tasks [15].

Future work will increase the number of instances the game is sampled over and the number of actions taken by each player to more rigorously investigate the impact of player decisions on the evolution of heterogeneous adversarial games. Further, there is merit in introducing additional populations of heterogeneous agents (i.e. more swarms) to study multi-agent teaming scenarios and the impact on conventional hierarchical C2 structures. Finally, there is scope to explore the application of multi-agent reinforcement learning [16, 17] to solve our complex cyber-physical systems through a game-theoretic lens.

## APPENDIX A – EXPERIMENTATION PARAMETERS

Below are the parameters used for the simulation.

*Table 1: Summary of parameters*

| Symbol | Value |
|---|---|
| $\sigma_1, \sigma_2$ | 0.5 |
| $\sigma_3$ | 8 |
| $\sigma_4$ | 4 |
| $\sigma_{12}, \sigma_{21}$ | 2 |
| $\sigma_{12_{Swarm}}, \sigma_{21_{Swarm}}$ | 0.5 |
| $\sigma_{12_{Headquarters}}, \sigma_{21_{Headqaurters}}$ | 5 |